\documentclass[%
 reprint,
 amsmath,amssymb,
 aps,
]{revtex4-2}

\usepackage{graphicx}
\usepackage{dcolumn}
\usepackage{bm}
\usepackage{multirow}

\usepackage[mathlines]{lineno}

\begin{document}

\preprint{APS/123-QED}

\title{The effect of initial nuclear deformation on dielectron photoproduction \protect\\ 
in hadronic heavy-ion  collisions}

\author{Jiaxuan Luo}

\author{Xinbai Li}

\author{Zebo Tang}%
 \author{Xin Wu}%

\author{Wangmei Zha}%
 \email{first@ustc.edu.cn}
 
 \affiliation{%
University of Science and Technology of China, Hefei 230026, China\\
}%




\date{\today}

\begin{abstract}
  Significant excesses of $e^+e^-$ pair production at very low transverse 
momentum ($p_T <$ 0.15 GeV/c) were observed by the STAR collaboration in hadronic heavy-ion collisions. Such enhancement is assumed to be a sign of photon-photon 
production in heavy-ion collisions with hadronic overlap, based on comparisons with model 
calculations for spherical Au + Au collisions. However, there is a lack of calculations for 
$e^+e^-$ pair production from coherent photon-photon interactions in hadronic U + U 
collisions, due to the deformity of Uranium nuclei. In this article, we present calculations for $e^+e^-$ pair photoproduction at 
$\sqrt{s_{NN}}$ = 193 GeV in both spherical and deformed U + U collisions within STAR detector
acceptance using the equivalent photon approximation (EPA). We 
conduct event-by-event analysis to investigate the effects of initial nuclear deformation on 
pair production. Our numerical results show good agreement with experimental data for the 
40\%--60\% and 60\%--80\% centrality classes in U + U collisions, and the differences between 
spherical and deformed configurations are approximately 3\%. We also calculate the yields of 
the photoproduced $e^+e^-$ pair in hadronic deformed Ru + Ru and Zr + Zr collisions at 
$\sqrt{s_{NN}}$ = 200 GeV. The results reveal that the ratios of the yields of Ru to Zr exhibit 
very small differences ($<$ 1\%) between spherical and deformed cases.

\end{abstract}

\maketitle


\section{\label{sec:level1}Introduction}
One of the major aims of the Relativistic Heavy Ion Collider (RHIC) at Brookhaven National 
laboratory (BNL) is to simulate the extreme conditions in the first microseconds of the 
Universe and search for the deconfined state of partonic matter, commonly 
known as quark-gluon plasma (QGP), in a laboratory \cite{ref1_Braun-Munzinger:2007edi, ref2_STAR:2005gfr}. 
Dileptons, which carry pure information about the hot and dense nuclear matter,
 are considered to be important probes for studying the properties 
of QGP since they are produced in the whole evolution of the collision and not involved 
in strong interactions \cite{ref3_Shuryak:1980tp}. Conventionally, dileptons are typically 
produced by the decays of known hadronic sources, QGP thermal radiation, and in-medium 
broadening of the $\rho$ spectral function \cite{ref4_vanHees:2007th, ref5_Rapp:2013nxa}.

In addition, dileptons can also be generated via photon-photon interactions in relativistic 
heavy-ion collisions \cite{ref6_Baltz:2007kq}. The almost transverse electromagnetic fields 
accompanied by the colliding nuclei can be viewed as an equivalent swarm of high-energy 
quasireal photons \cite{ref7_Bertulani:1987tz, ref8_Klein:2016yzr}. Emitted virtual photons 
from one nucleus can interact with those emitted by the other nucleus, leading to dilepton 
production, known as photoproduction process. Two-photon processes have been widely 
studied in ultraperipheral collisions (UPCs), where the impact parameter ($b$) is larger than 
twice the nuclear radius ($R_A$), and hadronic interactions do not occur 
\cite{ref9_STAR:2004bzo, ref10_PHENIX:2009xtn, ref11_ALICE:2013wjo}. 

Recently, significant 
enhancements of $e^+e^-$ pair production were observed by the STAR collaboration 
\cite{ref12_STAR:2018ldd} in peripheral Au + Au and U + U collisions ($b < 2R_A$). All 
detected excesses are found below $p_T \approx$ 0.15 GeV/c, whereas the upper limit of the 
transverse momentum of virtual photons is around 30 MeV/c ($k_{Tmax} \sim \hbar 
c/R_A$) \cite{ref13_Krauss:1997vr, ref14_Vidovic:1992ik}. Hence, this may indicate that the 
excesses likely result from photoproduction in violent hadronic heavy-ion collisions. 
Furthermore, model calculations of photon-photon interactions in spherical Au + Au 
collisions also support this idea \cite{ref15_Zha:2018ywo, ref16_Klein:2018cjh}. However, there are no calculations available on 
the photoproduced $e^+e^-$ pair in hadronic U + U collisions to date due to initial nuclear 
deformation.

The equivalent photon spectrum of a relativistic ion depends quadratically on its charge 
number $Z$ \cite{ref13_Krauss:1997vr}, and for this reason, the $e^+e^-$ pair produced 
by two-photon interactions should be proportional to $Z^4$. To further confirm that the 
excesses of the $e^+e^-$ pair at very low-$p_T$ are attributable to 
photon-photon processes, it is crucial to investigate the dependence of the observed 
excesses on the nuclear charge number. The isobaric collisions ($_{44}^{96}\textrm{Ru}$ + $_{44}^{96}\textrm{Ru}$ and 
$_{40}^{96}\textrm{Zr}$ + $_{40}^{96}\textrm{Zr}$) at $\sqrt{s_{NN}}$ = 200 GeV, proposed 
to search for the presence of the chiral magnetic effect \cite{ref18_STAR:2019bjg}, also 
provide a unique opportunity to verify the theory of photoproduction because similar 
hadronic backgrounds are expected due to the same nucleon number 
\cite{ref19_Voloshin:2010ut, ref20_STAR:2021mii}.

In this paper, we present the invariant mass dependence of the photoproduced $e^+e^-$
 pair for both spherical and deformed U + U collisions at $\sqrt{s_{NN}}$ = 193 GeV and 
compare our results with the excesses observed by STAR collaboration. We also calculate the 
$e^+e^-$ pair production in hadronic deformed Ru + Ru and Zr + Zr collisions at 
$\sqrt{s_{NN}}$ = 200 GeV, and the ratios of the yields of Ru to Zr between spherical and 
deformed cases are also shown.
\\

\section{Methodology}
\subsection{\label{sec:method1} Initial nuclear deformation}
The charge density for a spherical heavy ion is typically
 given by the Woods-Saxon distribution:
\begin{eqnarray}
  \rho_{sph}(r)=\frac{\rho_0}{1+e^{(r-R_0)/a}}
\label{eq1}
\end{eqnarray}
where $\rho_0$ represents the normalization factor and denotes the density at the center of 
the nucleus. The radius $R_0$ and skin depth $a$ are obtained from elastic electron scattering 
\cite{ref21_DeJager:1974liz, ref22_DeVries:1987atn}. However, for deformed nuclei, an 
alternative way to describe their charge density is to extend the two-parameter Fermi 
distribution by introducing deformation parameters:
\begin{eqnarray}
  \rho(\vec{r})=\frac{\rho_0}{1+{\rm exp}[\frac{r-R_0[1+\beta_2Y_{2}^{0}
  (\theta)+\beta_4Y_{4}^{0}(\theta)]}{a}]}
\label{eq2}
\end{eqnarray}
where $\beta_2$ and $\beta_4$ are quadrupole and hexadecupole deformation expressed in
the spherical-harmonics expansion, respectively \cite{ref23_Moller:1993ed}. It is noteworthy
 that this charge density is independent of the azimuthal angle, allowing us to derive 
$\rho(\vec{r}) = \rho(r, \theta)$.

Deformation parameters for the $_{92}^{238}\textrm{U}$ nucleus are taken from 
Ref.~[\onlinecite{ref24_Shou:2014eya}]. Nuclear density distributions are not clear for 
deformed $_{44}^{96}\textrm{Ru}$ and $_{40}^{96}\textrm{Zr}$ because e-A scattering 
experiments \cite{ref25_Raman:2001nnq, ref26_Pritychenko:2013taa} and comprehensive 
model deductions \cite{ref27_Moller:2008vcx} present significantly different results. In this 
study, we adopt larger $\beta_2$ values to evaluate the maximum impact of initial nuclear 
deformation on $e^+e^-$ pair photoproduction in hadronic heavy-ion collisions. The 
parameters for both spherical and deformed nuclei used in our analysis are listed in 
Table~\ref{tab1}. Additionally, the shape of the deformed nucleus is a prolate spheroid when 
$\beta_2 > 0$, and the direction of the major axis $\vec{v}$ in Eq.~(\ref{eq2}) is along the 
$z$ axis.

\begin{table}[b]
  \caption{\label{tab1}
  Woods-Saxon parameters for both spherical and deformed nuclei.
  }
  \begin{ruledtabular}
    \begin{tabular}{c|cc|cccc}
    \multirow{2}{*}{Nucleus} & \multicolumn{2}{c|}{Spherical} & \multicolumn{4}{c}{Deformed}                                        \\
                            & $R$ (fm)      & $a$ (fm)      & $R$ (fm)& $a$(fm)& $\beta_2$ &  $\beta_4$ \\\hline
    $_{92}^{238}\textrm{U}$  & 6.8054        & 0.605         & 6.8054 & 0.605   & 0.2863        & 0.093  \\
    $_{44}^{96}\textrm{Ru}$  & 5.085         & 0.46          & 5.085  & 0.46    & 0.158         & 0      \\
    $_{40}^{96}\textrm{Zr}$  & 5.02          & 0.46          & 5.02   & 0.46    & 0.217         & 0       
    \end{tabular}
  \end{ruledtabular}
\end{table}

In deformed heavy-ion collisions, the directions of the major axis of colliding nuclei 
$\vec{v}$ are expected to be random and irrelevant. Our calculations adopt the following 
reference frame: where the beam direction corresponds to the $z$ axis, and the direction of 
the impact parameter corresponds to the $x$ axis.

The charge density of a deformed nucleus with a specific $\vec{v}$ can then be expressed as:
\begin{equation}
  \rho_{\vec{v}}(\vec{r})=\rho[R_z^{-1}(-\varphi_v)R_y^{-1}(\theta_v)R_z^{-1}(\varphi_v)\vec{r}]
\label{eq3}
\end{equation}
\begin{equation}
  \vec{v}=({\rm sin}\theta_v{\rm cos}\varphi_v, {\rm sin}\theta_v{\rm sin}\varphi_v, {\rm cos}\theta_v)
\label{eq4}
\end{equation}
\begin{equation}
  R_y(\theta_v)=\begin{pmatrix}
    {\rm cos}\theta_v & 0 & {\rm sin}\theta_v \\ 
    0 & 1 & 0 \\ 
    -{\rm sin}\theta_v & 0 & {\rm cos}\theta_v 
    \end{pmatrix}
\label{eq5}
\end{equation}
\begin{equation}
  R_z(\varphi_v)=\begin{pmatrix}
    {\rm cos}\varphi_v & -{\rm sin}\varphi_v & 0 \\ 
    {\rm sin}\varphi_v & {\rm cos}\varphi_v & 0 \\ 
    0 & 0 & 1 
    \end{pmatrix}
\label{eq6}
\end{equation}
where $R_y(\theta_v)$ and $R_z(\varphi_v)$ are rotation matrices, and $\theta_v$ and 
$\varphi_v$ denote the polar angle and azimuthal angle of $\vec{v}$, respectively.
We assume that $\vec{v}$ is isotropic in the surface of the unit sphere, which means that 
cos$\theta_v$ is uniform in $[-1, 1]$ and $\varphi_v$ is uniform in $[0, 2\pi]$. In our 
calculations, the surface of the unit sphere is divided into 400 bins, leading to $N$ = 160000 
collision configurations when two deformed nuclei collide. Conventionally, configurations 
with $\vec{v_1} = \vec{v_2} = (0, 0, \pm 1)$ and $\vec{v_1} = \vec{v_2} = (\pm 1, 0, 0)$ are 
referred to as tip-tip and body-body collisions, respectively 
\cite{ref28_Haque:2011aa, ref29_Schenke:2014tga}, 
where subscripts 1 and 2 represent the two colliding nuclei. 
Selecting central tip-tip events and central body-body events based on experimental observables is 
possible \cite{ref30_Nepali:2007an}, so we will also present calculations for the two limiting 
cases in deformed U + U collisions.

\subsection{\label{sec:method2} Photon flux and form factor}
In relativistic heavy-ion collisions, the electric and magnetic fields accompanied by nuclei are 
mutually perpendicular and have the same absolute magnitudes. These almost transverse 
electromagnetic fields are very similar to the electromagnetic fields of photons and can be 
viewed as an equivalent swarm of quasireal photons \cite{ref8_Klein:2016yzr}. According to 
the equivalent photon approximation (EPA) method, the induced photon flux with energy 
$\omega$ at transverse position $\vec{x_{\perp }}$ from the center of the nucleus is given by 
\cite{ref7_Bertulani:1987tz}:
\begin{equation}
  n(\omega,\vec{x_{\perp }})=\frac{4Z^2\alpha}{\omega} \left | \int 
  \frac{{\rm d}^2 \vec{q_{\perp }}}{(2\pi)^2}\vec{q_{\perp }}\frac{F(\vec{q})}
  {\left | \vec{q}\right |^2 }e^{i\vec{x_{\perp }} \cdot \vec{q_{\perp }}}\right |^{2}
\label{eq7}
\end{equation}
\begin{equation}
  \vec{q}=(\vec{q_{\perp }},\frac{\omega}{\gamma})
\label{eq8}
\end{equation}
where $\alpha = 1/137$ is the fine-structure constant, $\gamma$ is the Lorentz factor of the 
nucleus, $Z$ is the nuclear charge number, and $\vec{q_{\perp }}$ is the transverse 
momentum of the photon. The form factor $F(q)$, carrying the information about the charge 
distribution inside the nucleus, can be obtained by performing a Fourier transformation to 
the charge density $\rho(\vec{r})$:
\begin{equation}
  F(\vec{q})=\int {\rm d}^3\vec{r}\rho(\vec{r})e^{i\vec{q}\cdot \vec{r}}
\label{eq9}
\end{equation}
For a spherical nucleus, the form factor can be expressed as follows:
\begin{equation}
  F(q)=\frac{4\pi}{q}\int {\rm d}r r\rho(r){\rm sin}(qr)
\label{eq10}
\end{equation}
For a spheroidal nucleus, the form factor depends on the direction of momentum
transfer as well:
\begin{eqnarray}
  F(q,\eta)=\iiint {\rm d}r{\rm d}\theta {\rm d}\varphi
   r^2{\rm sin}\theta\rho(r, \theta){\rm cos}[qr \nonumber \\
   \times ({\rm sin}\theta{\rm sin}\eta{\rm cos}\varphi %
  +{\rm cos}\theta{\rm cos}\eta)]
\label{eq11}
\end{eqnarray}
\begin{equation}
  {\rm cos}\eta=\frac{\vec{q}\cdot\vec{v}}{\left | \vec{q}\right |}
\label{eq12}
\end{equation}
where $\eta $ denotes the angle between momentum transfer $\vec{q}$ and major axis 
$\vec{v}$. Utilizing Eq.~(\ref{eq7}), we can calculate the photon flux 
$n_{\vec{v}}(\omega,\vec{x_{\perp }})$ for a deformed nucleus with a given $\vec{v}$.

\begin{figure}[t]
  \includegraphics[width=0.95\linewidth]{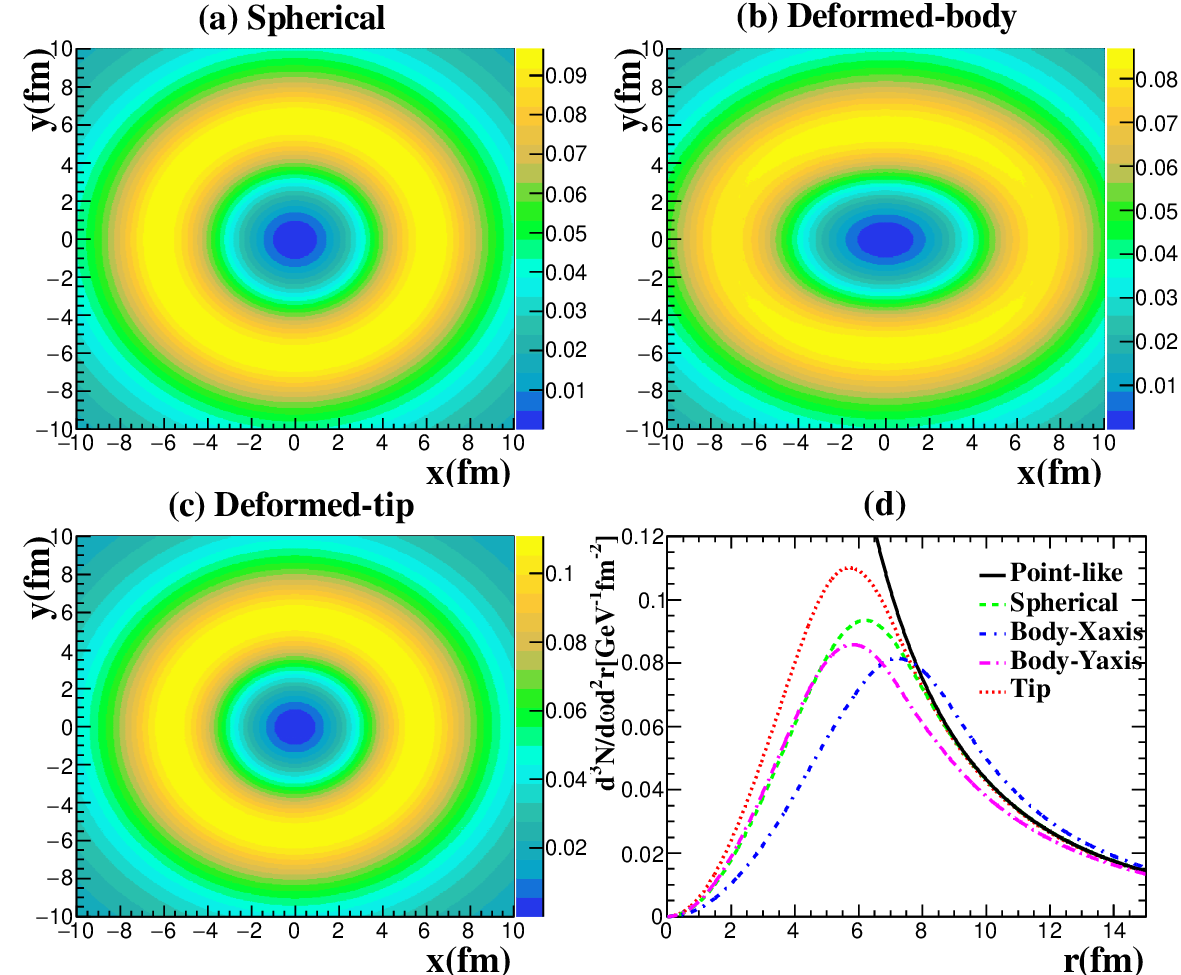}
  \caption{\label{label_fig1} 
  The photon flux distributions with energy $\omega$ = 1 GeV in U + U collisions at 
  $\sqrt{s_{NN}}$ = 193 GeV as a function of transverse position $\vec{x_{\perp }}$ from the 
  center of the nucleus. Panel (a): spherical nucleus, panels (b) and (c): body and tip orientations 
  for deformed nucleus, and panel (d): one-dimensional photon flux from different 
  configurations as a function of distance $r$ from the center of nucleus. The solid line 
  represents the photon flux with a point-like form factor.  
  }
\end{figure}

Fig.~\ref{label_fig1} shows the photon flux distributions with energy $\omega$ = 1 GeV in U 
+ U collisions at $\sqrt{s_{NN}}$ = 193 GeV as a function of transverse position 
$\vec{x_{\perp }}$ from the center of the nucleus. The photon flux for the spherical nucleus is 
shown in panel (a), and those in the case of body and tip orientations for the deformed 
nucleus are presented in panels (b) and (c). The results from different configurations as a 
function of distance $r$ from the center of the nucleus are illustrated in panel (d), and one 
can observe that the differences are concentrated around $R_0$. The photon flux from the 
tip orientation is greater than that for the spherical nucleus, while the maximum region 
(orange circular band) presents a smaller radius. The pattern from the body orientation 
exhibits an ellipse, where the extreme points of photon flux along the $x$-axis and $y$-axis 
differ, corresponding to the polar (major) radius and equatorial radius of the prolate spheroid, 
respectively.

\subsection{\label{sec:method3}  $e^+e^-$ pair photoproduction}
According to the equivalent photon approximation, the cross section of the $e^+e^-$ pair 
produced by the two-photon process in relativistic heavy-ion collisions can be expressed as 
\cite{ref13_Krauss:1997vr}:
\begin{eqnarray}
  \label{eq13}
  \sigma(AA \rightarrow AA e^+ e^-)=\int {\rm d}\omega_1\int {\rm d}\omega_1 
  n_1(\omega_1)n_2(\omega_2)\nonumber \\
  \times\sigma(\gamma \gamma \rightarrow e^+ e^-)
\end{eqnarray}
where $\sigma(\gamma \gamma \rightarrow e^+ e^-)$ is the photon-photon reaction cross 
section for the $e^+e^-$ pair. The energy of the produced particles is $E = 
\omega_{1}+\omega_{2}$, while their longitudinal momentum becomes $p_z=\omega_{1}- 
\omega_{2}$ as the velocity of the moving heavy ion approaches the speed of light. The final-
state particles have a small transverse momentum, which can be negligible compared to 
longitudinal momentum. Consequently, the invariant mass $W$ and rapidity $y$ of the 
$e^+e^-$ pair can be obtained as follows:
\begin{equation}
  \label{eq14}
  W=\sqrt{E^2-p^2}=\sqrt{4\omega_1\omega_2}
\end{equation}
\begin{equation}
  \label{eq15}
  y=\frac{1}{2}{\rm ln}\frac{E+p_z}{E-p_z}=\frac{1}{2}{\rm ln}\frac{\omega_1}{\omega_2}
\end{equation}
Therefore, ${\rm d}\omega_1{\rm d}\omega_2$ in Eq.~(\ref{eq13}) can be converted to ${\rm 
d}W{\rm d}y$. The cross section for producing a pair of electrons with invariant mass $W$ is 
given by the Breit-Wheeler formula \cite{ref31_Brodsky:1971ud}:
\begin{eqnarray}
  \label{eq16}
  \sigma(\gamma \gamma \rightarrow e^+ e^-)=\frac{4\pi^2\alpha^2}{W^2}
  [(2+\frac{8m_e^2}{W^2}-\frac{16m_e^4}{W^4}) \nonumber \\
  \times{\rm ln}(\frac{W+\sqrt{W^2-4m_e^2}}{2m_e})-\sqrt{1-\frac{4m_e^2}{W^2}}
  (1+\frac{4m_e^2}{W^2})]
\end{eqnarray}
where $m_e$ is the mass of the electron.
\begin{table}[hbtp]
  \caption{\label{tab2}
  Centrality definition for both spherical and deformed U + U collisions.
  }
  \begin{ruledtabular}
  \begin{tabular}{c|c|ccc}
  Centrality                 & Configuration & \multicolumn{1}{c}{ $b$ range(fm)} & \multicolumn{1}{c}{$N_{part}$ range} & \multicolumn{1}{c}{$\left \langle N_{part}\right \rangle$} \\\hline
  \multirow{4}{*}{40\%--60\%} & Spherical     & 10.4--12.7                       & 29.7--93.8                         & 57.6                          \\
                             & Tip-tip       & 9.8--12.0                        & 28.5--93.1                         & 57.6                          \\
                             & Body-body     & 11.8--14.4                       & 35.1--102.8                        & 65.1                          \\
                             & Deformed      &                                 & 30.1--94.2                         & 58.0                           \\\hline
  \multirow{4}{*}{60\%--80\%} & Spherical     & 12.7--14.7                       & 5.9--29.7                          & 15.3                          \\
                             & Tip-tip       & 12.0--13.8                       & 5.6--28.5                          & 15.2                          \\
                             & Body-body     & 14.4--16.7                       & 7.1--35.1                          & 18.4                          \\
                             & Deformed      &                                 & 6.0--30.1                          & 15.6                         
  \end{tabular}
  \end{ruledtabular}
\end{table}
\begin{table*}[!hbtp]
  \caption{\label{tab3}
  Centrality definition for both spherical and deformed Ru + Ru collisions.
  }
  \begin{ruledtabular}
    \begin{tabular}{c|ccc|cc}
      \multirow{2}{*}{Centrality} & \multicolumn{3}{c|}{Spherical}                                                                        & \multicolumn{2}{c}{Deformed}                                            \\
                                  & \multicolumn{1}{c}{$b$ range (fm)} & \multicolumn{1}{c}{$N_{part}$ range} & \multicolumn{1}{c|}{$\left \langle N_{part}\right \rangle$} & \multicolumn{1}{c}{$N_{part}$ range} & \multicolumn{1}{c}{$\left \langle N_{part}\right \rangle$} \\ \hline
      0--10\%                      & 0--3.7                            & 128.0--180.9                       & 152.7                         & 127.6--180.1                       & 152.0                                \\
      10\%--20\%                   & 3.7--5.2                          & 90.8--128.0                        & 108.4                         & 90.7--127.6                        & 108.2                               \\
      20\%--30\%                   & 5.2--6.3                          & 63.3--90.8                         & 76.4                          & 63.2--90.7                         & 76.3                                \\
      30\%--40\%                   & 6.3--7.3                          & 42.5--63.3                         & 52.3                          & 42.5--63.2                         & 52.3                                \\
      40\%--50\%                   & 7.3--8.2                          & 27.1--42.5                         & 34.3                          & 27.1--42.5                         & 34.3                                \\
      50\%--60\%                   & 8.2--9.0                          & 16.1--27.1                         & 21.2                          & 16.1--27.1                         & 21.2                                \\
      60\%--70\%                   & 9.0--9.7                          & 8.7--16.1                          & 12.1                          & 8.8--16.1                          & 12.1                                \\
      70\%--80\%                   & 9.7--10.4                         & 4.3--8.7                           & 6.3                           & 4.3--8.8                           & 6.3                                 \\
      80\%--90\%                   & 10.4--11.1                        & 1.7--4.3                           & 2.9                           & 1.7--4.3                           & 2.9                                
    \end{tabular}
  \end{ruledtabular}
\end{table*}
\begin{table*}[!hbtp]
  \caption{\label{tab4}
  Centrality definition for both spherical and deformed Zr + Zr collisions.
  }
  \begin{ruledtabular}
    \begin{tabular}{c|ccc|cc}
      \multirow{2}{*}{Centrality} & \multicolumn{3}{c|}{Spherical}                                                                        & \multicolumn{2}{c}{Deformed}                                            \\
                                  & \multicolumn{1}{c}{$b$ range (fm)} & \multicolumn{1}{c}{$N_{part}$ range} & \multicolumn{1}{c|}{$\left \langle N_{part}\right \rangle$} & \multicolumn{1}{c}{$N_{part}$ range} & \multicolumn{1}{c}{$\left \langle N_{part}\right \rangle$} \\ \hline
      0--10\%                      & 0--3.6        & 128.3--181.1   & 153.0     & 127.6--179.5   & 151.7           \\
      10\%--20\%                   & 3.6--5.1      & 91.0--128.3    & 108.7     & 90.7--127.6    & 108.2           \\
      20\%--30\%                   & 5.1--6.3      & 63.3--91.0     & 76.5      & 63.2--90.7     & 76.4            \\
      30\%--40\%                   & 6.3--7.3      & 42.4--63.3     & 52.4      & 42.5--63.2     & 52.4            \\
      40\%--50\%                   & 7.3--8.1      & 27.0--42.4     & 34.3      & 27.1--42.5     & 34.3            \\
      50\%--60\%                   & 8.1--8.9      & 16.0--27.0     & 21.1      & 16.1--27.1     & 21.2            \\
      60\%--70\%                   & 8.9--9.6      & 8.7--16.0      & 12.0      & 8.7--16.1      & 12.1            \\
      70\%--80\%                   & 9.6--10.3     & 4.3--8.7       & 6.2       & 4.2--8.7       & 6.3             \\
      80\%--90\%                   & 10.3--11.0    & 1.7--4.3       & 2.8       & 1.7--4.3       & 2.9     
    \end{tabular}
  \end{ruledtabular}
\end{table*}

The model calculations of $e^+e^-$ pair photoproduction have been presented in hadronic 
Au + Au collisions \cite{ref15_Zha:2018ywo}, and we utilize a similar approach to conduct 
model calculations of $e^+e^-$ pair photoproduction in randomly oriented collisions of 
deformed heavy ions. The yield for the photoproduced $e^+e^-$ pair with the orientation 
$(\vec{v_1}, \vec{v_2})$ in a selected centrality bin can be expressed as:
\begin{widetext}
\begin{equation}
  \label{eq17}
  \frac{{\rm d}^2N}{{\rm d}W{\rm d}y}=\frac{\frac{W}{2}\int_{b_{min}}^{b_{max}}{\rm d}^2
  \vec{b}\int {\rm d}^2\vec{x_{\perp} }n_{\vec{v_1}}(\omega_1,\vec{x_{\perp }})
  n_{\vec{v_2}}(\omega_2,\vec{x_{\perp }}-\vec{b})\sigma(\gamma \gamma \rightarrow e^+ e^-)
  P_{\vec{v_1},\vec{v_2}}^B(\vec{b})}{\int_{b_{min}}^{b_{max}}{\rm d}^2\vec{b}P_{\vec{v_1},\vec{v_2}}^B(\vec{b})}
\end{equation}
\end{widetext}
where $b_{min}$ and $b_{max}$ are the minimum and maximum impact parameters for a
 given centrality class, and $P_{\vec{v1},\vec{v2}}^B(\vec{b})$ is the probability of hadronic 
interactions:
\begin{equation}
  \label{eq18}
  P_{\vec{v_1},\vec{v_2}}^B(\vec{b})=1-{\rm exp}[-A^2\sigma_{NN} 
  \int {\rm d}^2\vec{s}
  T_{\vec{v_1}}(\vec{s})T_{\vec{v_2}}(\vec{s}-\vec{b})]
\end{equation}
where $A$ is the nucleon number, $\sigma_{NN}$ is the inelastic nucleon-nucleon cross 
section, which is dependent on collision energy 
$\sqrt{s_{NN}}$ \cite{ref32_dEnterria:2020dwq}, and the nuclear thickness function 
$T_{\vec{v}}(\vec{s})$ is the projection of nuclear charge density with orientation $\vec{v}$ on 
the $x$-$y$ plane:
\begin{equation}
  \label{eq19}
  T_{\vec{v}}(\vec{s})=\int {\rm d}z \rho_{\vec{v}}(\vec{s},z)
\end{equation}

In this way, we can directly obtain the $e^+e^-$ pair yields in tip-tip and body-body 
collisions, but the calculations of all $N$ = 160000 collision configurations are required to 
obtain the average yields in deformed heavy-ion collisions. Our results are filtered to match 
the fiducial acceptance ($p_T^e >$ 0.2 GeV/c, $ \left | \eta^e \right | < $ 1, $ \left | y^{ee} 
\right | < $ 1) to compare with experimental data from the STAR collaboration 
\cite{ref12_STAR:2018ldd}. As discussed in Refs.~\cite{ref15_Zha:2018ywo, 
ref33_Zha:2017jch}, the impact of violent hadronic interactions occurring in the overlap region 
on photoproduction is negligible for peripheral collisions. In central collisions, this effect on 
differences between spherical and deformed configurations should be small. Therefore, we 
neglect the possible disruptive effects from hadronic interactions in our calculations.

 \subsection{\label{sec:method4}  Centrality definition}
 \begin{figure*}[!htbp]
  \includegraphics[width=1.0\linewidth]{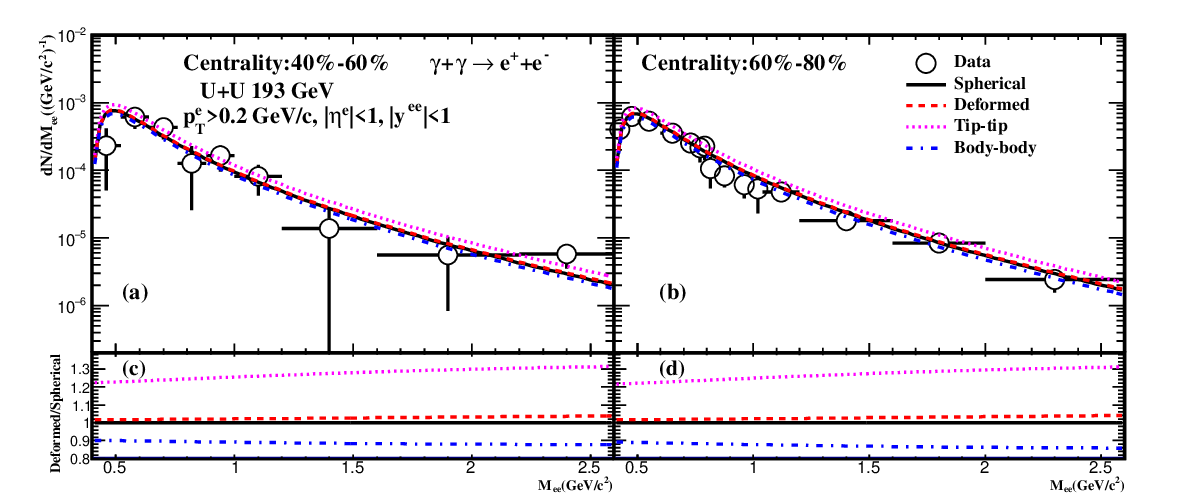}
  
  \caption{\label{label_fig2} 
  The $e^+e^-$ pair mass spectra ${\rm d}N/{\rm d}M$ within the STAR acceptance 
in (a) 40\%--60\% and (b) 60\%--80\% for spherical, deformed, tip-tip and body-body U + U 
collisions at $\sqrt{s_{NN}}$ = 193 GeV, compared with experimental data from the STAR 
collaboration \cite{ref12_STAR:2018ldd}. Panels (c) and (d) present the ratios of $e^+e^-
$ pair mass spectra from different configurations to those from the spherical case.
  }
\end{figure*}

In deformed heavy-ion collisions, we will employ the Glauber model 
\cite{ref32_dEnterria:2020dwq, ref34_Miller:2007ri} to define centrality and provide 
corresponding impact parameters. For a random collision configuration with the orientation 
$(\vec{v_1}, \vec{v_2})$, the centrality can be expressed as a percentage of the interaction 
probability:
\begin{equation}
  \label{eq20}
  c_i(b)=\frac{\int_{0}^{b}{\rm d}^2\vec{{b}'}P_{\vec{v_1},\vec{v_2}}^B(\vec{{b}'}) }
  {\int_{0}^{\infty}{\rm d}^2\vec{{b}'}P_{\vec{v_1},\vec{v_2}}^B(\vec{{b}'})}
\end{equation}
In tip-tip and body-body collisions, this approach is suitable, but it is not sufficient when 
calculating average yields because all configurations occur with the same probability. Instead, 
the two-component approach $fN_{coll} + (1-f)N_{part}$ is a better choice 
\cite{ref34_Miller:2007ri, ref35_Masui:2009qk, ref36_Kharzeev:2000ph},
where $N_{part}$ is the number of participating nucleons, and $N_{coll}$ is the number of 
nucleon-nucleon collisions \cite{ref34_Miller:2007ri}: 
\begin{flalign}
  \label{eq21}
  N_{part}(b)&=A\int {\rm d}^2\vec{s}T_{\vec{v_1}}(\vec{s})
  \left \{1-[1-T_{\vec{v_2}}(\vec{s}-\vec{b})\sigma_{NN}]^A\right \}  \nonumber \\
  &+A\int {\rm d}^2\vec{s}T_{\vec{v_2}}(\vec{s}-\vec{b})
  \left \{1-[1-T_{\vec{v_1}}(\vec{s})\sigma_{NN}]^A\right \}
\end{flalign}

  
\begin{equation}
  \label{eq22}
  N_{coll}(b)=A^2\sigma_{NN} \int {\rm d}^2\vec{s}
  T_{\vec{v_1}}(\vec{s})T_{\vec{v_2}}(\vec{s}-\vec{b})
\end{equation}
The relative weight $f$ is usually small
 \cite{ref32_dEnterria:2020dwq, ref35_Masui:2009qk, ref36_Kharzeev:2000ph, ref37_ALICE:2013hur, ref38_PHOBOS:2004hlv}, and thus we set $f$ = 
0 for simplicity. Therefore, the centrality in deformed heavy-ion collisions is defined by the 
cumulative distribution function of $N_{part}$:
\begin{equation}
  \label{eq23}
  c= \int_{N_{part}}^{\infty }{\rm d}{N}'_{part}P({N}'_{part})
\end{equation}
\begin{equation}
  \label{eq24}
  P(N_{part})=\frac{\sum_{i=1}^{N}P_i(N_{part})}{N}
\end{equation}
where $P(N_{part})$ is the average probability distribution of $N_{part}$ and $P_i(N_{part})$ is 
the probability distribution for a special configuration, which can be calculated using 
Eqs.~(\ref{eq20}) and (\ref{eq21}):
\begin{equation}
  \label{eq25}
  P_i(N_{part})=-\frac{{\rm d}c_i(b)}{{\rm d}N_{part}(b)}
\end{equation}

It is noteworthy that $N_{part}(b)$ monotonically decreases with impact parameter $b$. Once 
the range of $N_{part}$ in a given centrality class is obtained from Eq.~(\ref{eq23}), the 
corresponding range of the impact parameter for a random configuration can be determined 
from Eq.~(\ref{eq21}). Then, the yield for the photoproduced $e^+e^-$ pair can be 
calculated using Eq.~(\ref{eq17}). Table~\ref{tab2} presents the centrality definitions in 40\%--60\% and 60\%--80\% for U + U collisions as well as the tip-tip and body-body configurations. 
The average number of participants $\left \langle N_{part}\right \rangle$ is also listed in the 
table. Tables~\ref{tab3} and \ref{tab4} report the centrality definitions for Ru + Ru and Zr + 
Zr collisions, respectively, under both spherical and deformed configurations.
Despite the systematic differences of $\left \langle N_{part}\right \rangle$ observed when compared with the
Glauber Monte Carlo approach \cite{ref34_Miller:2007ri}, the variation in impact parameter between the two
calculations is found to be minor.

 \begin{figure*}[!hbtp]
  \includegraphics[width=1.0\linewidth]{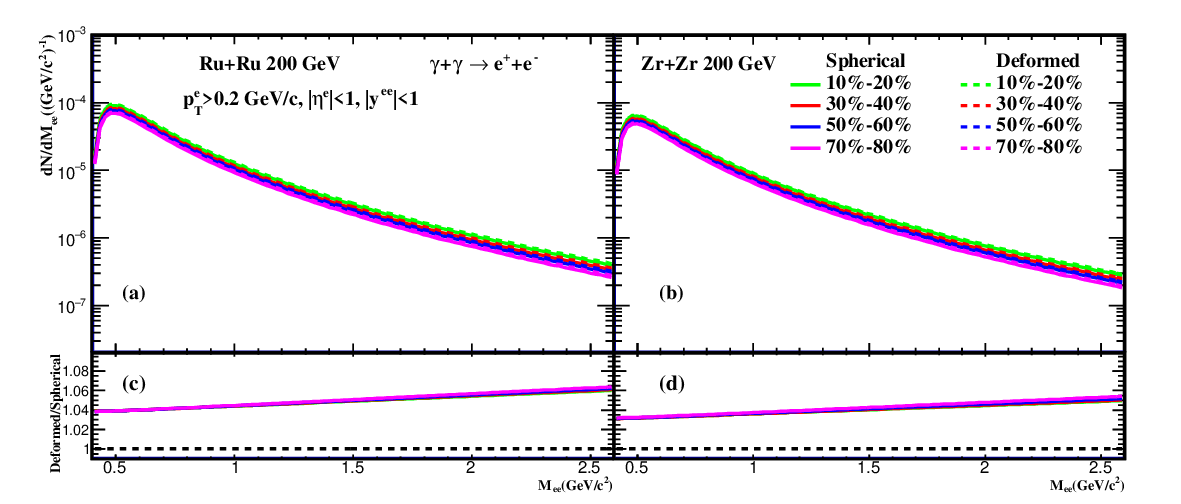}
  
  \caption{\label{label_fig3} 
  The $e^+e^-$ pair mass spectra ${\rm d}N/{\rm d}M$ in (a) Ru + Ru and (b) Zr + 
Zr collisions at $\sqrt{s_{NN}}$ = 200 GeV for different centrality classes within STAR 
acceptance. The solid and dashed lines represent spherical and deformed configurations, 
respectively. Panels (c) and (d) present the ratios of $e^+e^-$ pair mass spectra in deformed 
collisions to those in spherical collisions.
  }
\end{figure*}

\section{Results}

Fig.~\ref{label_fig2} shows the $e^+e^-$ pair mass spectra ${\rm d}N/{\rm d}M$ within the 
STAR acceptance in (a) 40\%--60\% and (b) 60\%--80\% for spherical, tip-tip and body-body 
U + U collisions at $\sqrt{s_{NN}}$ = 193 GeV. The spectra are contrasted with previously 
reported excess yields at low $p_T$ from the STAR collaboration \cite{ref12_STAR:2018ldd}, 
while the average yields in deformed heavy-ion collisions are also presented and denoted as 
“Deformed”. The ratios of $e^+e^-$ pair mass spectra from different configurations to those 
from the spherical case are shown in panels (c) and (d). The result in tip-tip collisions is 
approximately 25\% higher than that in spherical collisions. The difference becomes more 
significant as the invariant mass $M_ee$ increases due to the higher energy of photons 
induced in tip-tip collisions. Meanwhile, the pair mass spectrum in body-body collisions is 
approximately 10\% lower than that in spherical collisions. Both the deformed and spherical 
configurations can describe the data well, and the former is slightly higher by approximately 
3\% than the latter.

Fig.~\ref{label_fig3} shows the $e^+e^-$ pair mass spectra ${\rm d}N/{\rm d}M$ in (a) Ru + 
Ru and (b) Zr + Zr collisions at $\sqrt{s_{NN}}$ = 200 GeV for different centrality classes within 
the STAR acceptance. Likewise, the ratios of $e^+e^-$ pair mass spectra in deformed 
collisions to those in spherical collisions are shown in panels (c) and (d). Compared to the Zr 
+ Zr collisions, the $e^+e^-$ pair yields for Ru + Ru collisions are higher due to the larger 
charge number. The pair mass spectra with the deformed configuration exhibit approximately 
5\% increases compared to the spherical case in Ru + Ru collisions, while the differences 
become slightly smaller in Zr + Zr collisions. Although the yields of $e^+e^-$ pair yields increase in more central collisions, the ratios do not seem to exhibit dependence on centrality.

\begin{figure}[!hbtp]
  \includegraphics[width=0.95\linewidth]{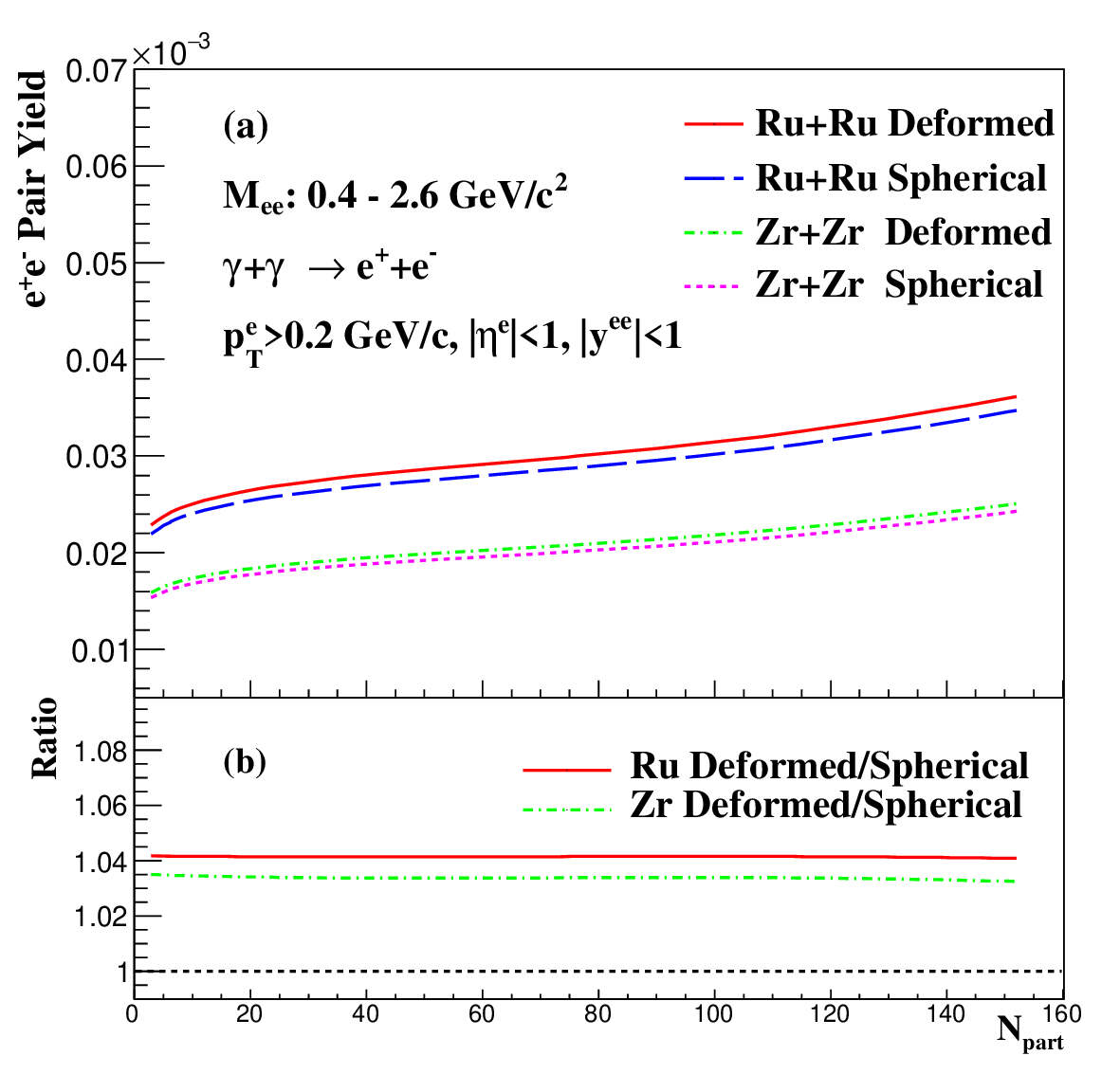}
  \caption{\label{label_fig4} 
  (a) The integrated yields of the photoproduced $e^+e^-$ pair as a function of 
  $N_{part}$ in the mass region of 0.4--2.6 GeV/$c^2$ in Ru + Ru and Zr + Zr collisions with 
  both spherical and deformed configurations. (b) The corresponding ratios of $e^+e^-$ pair 
  yields in deformed collisions to those in spherical collisions.  
  }
\end{figure}

We further present the centrality dependence of integrated yields of photoproduced 
$e^+e^-$ pair in the mass region of 0.4--2.6 GeV/$c^2$ in Ru + Ru and Zr + Zr collisions 
with both spherical and deformed configurations in Fig.~\ref{label_fig4}. The corresponding 
ratios of $e^+e^-$ pair yields in deformed collisions to those in spherical collisions are shown 
in panel (b), and indeed, the impact of initial nuclear deformation on $e^+e^-$ pair 
photoproduction does not have centrality dependence.

\begin{figure}[!hbtp]
  \includegraphics[width=0.94\linewidth]{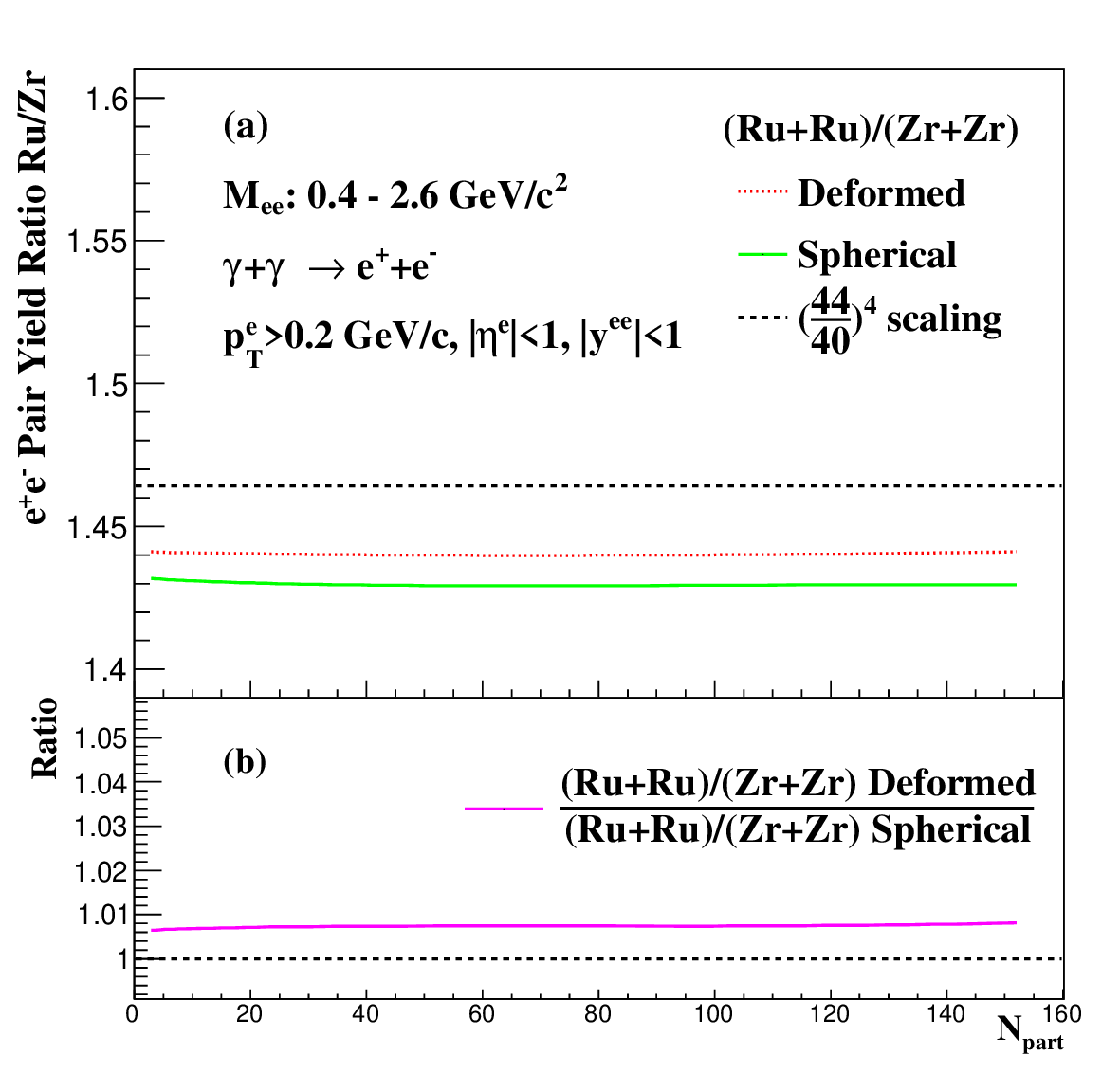}
  \caption{\label{label_fig5} 
  (a) The ratios of $e^+e^-$ pair yields in Ru + Ru collisions to those in Zr + Zr 
collisions as a function of $N_{part}$ in the mass region of 0.4--2.6 GeV/$c^2$. The solid line 
represents the spherical configuration, while the dotted line represents the deformed 
configuration. (b) The difference between the two ratios.
  }
\end{figure}

Fig.~\ref{label_fig5} illustrates the ratios of $e^+e^-$ pair yields in Ru + Ru collisions to those 
in Zr + Zr collisions as a function of $N_{part}$. The ratios are slightly smaller than the $(\frac{44}{40})^4$ scaling, which is due to the slightly different Woods-Saxon parameters for Zr and Ru nuclei. And one can observe that the difference between the two ratios for spherical and deformed configurations 
is very small ($<$ 1\%), which demonstrates that the impact of initial nuclear deformation on the ratios of 
$e^+e^-$ pair photoproduction between Ru + Ru and Zr + Zr collisions is negligible.
\section{Summary}
In summary, we employ the Glauber model and equivalent photon approximation to 
investigate the impact of initial nuclear deformation on $e^+e^-$ pair photoproduction in 
hadronic U + U, Ru + Ru and Zr + Zr collisions. In this study, we present the equivalent photon 
flux distributions as a function of transverse position for deformed colliding nuclei with a 
random orientation. We conduct calculations of $e^+e^-$ pair photoproduction in hadronic 
heavy-ion collisions considering both spherical and deformed configurations. Our results can 
describe the experimental data well for 40\%--60\% and 60\%--80\% centrality classes in U + 
U collisions. We also observe approximately 3\% differences between spherical and deformed 
configurations. The impact of initial nuclear deformation on the ratios of $e^+e^-$ pair 
photoproduction between Ru + Ru and Zr + Zr collisions is negligible ($<$ 1\%). This 
observation may alleviate difficulties for future study of $e^+e^-$ pair photoproduction in 
isobaric collisions.\\
\\
\section{Acknowledgments}
This work is supported in part by the National Key Research and Development Program of China under 
Contract No. 2022YFA1604900 and the National Natural Science Foundation of China (NSFC) under Contract
No. 12175223 and 12005220. W. Zha is supported by Anhui Provincial Natural Science Foundation 
No. 2208085J23 and Youth Innovation Promotion Association of Chinese Academy of Sciences.
\bibliography{ref}
\end{document}